\documentclass[aps,twocolumn,groupedaddress]{revtex4}
\usepackage{graphicx}
\begin{document}
\title{Absolute Hydration Free Energies of Ions, Ion-Water Clusters, and Quasi-chemical Theory}
\author{D. Asthagiri}
\author{Lawrence R. Pratt}
\author{H. S. Ashbaugh}
\affiliation{Theoretical Division, Los Alamos National Laboratory, Los
Alamos NM 87545} 
\date{\today}
\begin{abstract}
Experimental studies on ion-water clusters have provided insights into the
microscopic aspects of hydration phenomena. One common view is that
 extending those experimental studies to larger cluster sizes
 would give the single ion absolute hydration free energies not
 obtainable by classical thermodynamic methods. This issue is reanalyzed
 in the context of recent computations and molecular theories on ion
 hydration, particularly considering the hydration of H$^+$, Li$^+$,
 Na$^+$, and HO$^-$ ions and thence the hydration of neutral ion
 pairs. The hydration free energies of neutral pairs computed here are
 in good agreement with experimental results, whereas the calculated
 absolute hydration free energies, and the excess chemical potentials,
 deviate consistently from some recently tabulated hydration free
 energies based on ion-water cluster data. We show how the single ion
 absolute hydration free energies are not separated from the potential
 of the phase in recent analyses of ion-water cluster data, even in
 the limit of  large cluster sizes.  We conclude that naive
 calculations on ion-water clusters ought to agree with results
 obtained from experimental studies of ion-water clusters because both
 values include the contribution, somewhat extraneous to the local
 environment of the ion, from the potential of the phase.
 \end{abstract}

\maketitle

\section{Introduction}

The interactions between the ions and inner shell water molecules are
much stronger than thermal energies and typically display chemical
complexities even in the simplest cases \cite{lrp:jacs00,Pratt:ES:99}.
Computational chemistry and the theories of solutions are
approaching the stage that calculations on the free energies of ions
in water must be taken seriously.  Despite this progress, there is
still a non-trivial level of confusion about what is to be learned
from experimental comparisons, particularly with comparisons developed
from current experimental studies of ion-water clusters.  This paper
attempts to provide guidance on these issues by presenting a thermodynamic
 reconsideration of conclusions drawn from ion-water cluster experiments, and 
 results of recent molecular calculations.

Computational results are presented for aqueous ions H$^+$, Li$^+$,
Na$^+$, and HO$^-$ which participate in a myriad of biological
processes \cite{alberts:02}. These are common ions in aqueous phase
chemistry, and this is especially true for H$^+$ and HO$^-$.  These 
species also play a significant role in current
problems such as the speciation of beryllium metal in the environment
\cite{sauer:02} or in the lungs.  Understanding beryllium speciation,
and the development of beryllium toxicity, is of substantial
technological interest, and would benefit from a molecular
understanding of ion hydration phenomena.

Absolute hydration free energies, or equivalently single ion
activities, that we wish to clarify are not measureable by {\em
purely} thermodynamic means \cite{guggenheim:29,guggenheim:67};
correct single ion activities will contain fundamental
extra-thermodynamical information and hence their usefulness in
thermodynamical analysis might be questioned.
Extra-thermodynamical, single ion activities, however, are not unknowable (see
\cite{oppenheim:jpc64} and \cite{lrp:jpc92}).  Furthermore, that
knowledge would clarify the molecular understanding of ion hydration,
and permit a compact, tabulated consensus of molecularly valid,
experimental thermodynamic information.

The existing tables, for example
\cite{Friedman:73,CONWAYBE:EVAUPI,MARCUSY:SIMEDT}, should naturally be
consistent in describing hydration free energies for neutral
combinations of ions, since those combinations are thermodynamically
measurable.  But the alignment of those tables depends on
extra-thermodynamical information.  An accurate determination of the
properly defined single ion activity for any ion represented in those
tables would be sufficient to align those tables.  It has been pointed
out recently \cite{lrp:jpca02} that in favorable cases the 
inaccuracies of computed single ion hydration free energies could
probably be made less than the misalignments in current tables of
single ion hydration free energies.  The present paper pursues this
possibility further.

Because simulated extra-thermodynamical information is available from
molecular simulations,  it is possible to compute solely the work of
coupling the ion with the solvent in its locality, {\em i.e.,\/}  the absolute 
 hydration free energies
of the ion separated from the contribution from the potential of
the phase (section II).  Simulation results, however, critically depend on
parameterized molecular models and on the treatment of electrostatic
interactions. Moreover, for many ions, particularly H$^+$ and
HO$^-$, the interactions with water are not easily described by
convenient force-fields. This problem is circumvented by the
quasi-chemical theory of solutions, within which the
excess chemical potential of the solute is partitioned into an
inner-sphere contribution, accounting for the chemistry in the inner
shell, and an outer-sphere contribution. This latter contribution    
describes the interaction of the inner-sphere quasi-component with the
rest of the fluid and can usually be well described by simple 
force-field models or dielectric continuum models. This quasi-chemical
approach has been used before to treat ions, such as, H$^+$
\cite{lrp:jpca02}, Li$^+$\cite{lrp:jacs00}, Na$^+$\cite{lrp:fpe01},
and HO$^-$\cite{asthagir:02}. Chemical reactions in solution, such as
speciation of Fe$^{3+}$\cite{rlmartin:jpc98} and
Be$^{2+}$\cite{asthagir:be02}, are also well described. 

In this article we bring together our earlier observations on the
hydration of monovalent ions in conjunction with classical force-field
based estimates of outer-sphere contributions. The results so 
obtained for neutral ion pairs are in good accord with
experiments, but our single ion values differ consistently
from current tabulations based on cluster-ion hydration experiments. 

Our plan for this report is as follows: First, we define our problem
and establish the notation we will use.  We then discuss
the experimental studies leading to current estimates of absolute
hydration free energies of the ions. After that, we present the
quasi-chemical theory and our computational results. Finally, we
return to analyze the experimental cluster-ion hydration results in
this context.  In the last section we identify conclusions of this
work.

\section{Absolute hydration free energies of ions in water}

We will consider the chemical potentials of ionic solutes in liquid
water and we cast these chemical potentials in the form
\begin{eqnarray}
\mu^{}_{\mathrm{M}^{+}}(w) & = & e\phi (w) \nonumber \\
& + & RT \ln \left\lbrack\rho_{\mathrm{M}^{+}} (w) \Lambda_{\mathrm{M}^{+}}{}^3\right\rbrack \nonumber \\
& + &\overline \mu_{\mathrm{M}^{+}} (w)~.
\label{chempot}
\end{eqnarray}
This treats an ionic solute M$^{+}$ suggested to be a metal ion of
charge $e$, but the  notation  here will be extended naturally to anions
X$^{-}$.  The qualifier $(w)$ in Eq.~\ref{chempot} indicates the
macroscopic phase to which this chemical potential, or another
quantity there, applies.  Thus, $\rho_{\mathrm{M}^{+}} (w)$ is the
number density of M$^{+}$ in a liquid water phase.  (Though we will be
particularly interested in the case of liquid water as a solvent,
other cases must be permitted also in the typical thermodynamic
analysis.)  The quantity $\phi (w)$ is the electrostatic potential of
the $w$ phase.  This may be introduced on the basis of the principle
that a mean electric field must be zero through interior regions of a
macroscopic conductor at equilibrium, and we will use the language that $\phi (w)$ is
the \emph{potential of the phase.}  Only differences in these
potentials between phases will be involved, however, in our
analysis.  The quantity $\Lambda_{\mathrm{M}^{+}}$ is the thermal
deBroglie wavelength for the species M$^{+}$, a known function of the
internal characteristics of an M$^{+}$ ion, and of the temperature
$T$, but with no dependences of further thermodynamic relevence.  (The
standard state adopted by Eq.~\ref{chempot} is sometimes referred to
as the Ben-Naim standard state.)

Finally, $\overline\mu_{\mathrm{M}^{+}} (w)$ is the object of the
present study.  It depends on temperature, pressure, and
composition of the system.  In the limit of vanishing solute
concentration we will call it the \emph{absolute hydration free
energy} of M$^{+}$.  We are principally interested in standard
temperature and pressure conditions.  As Eq.~\ref{chempot} is
formulated, $\overline\mu_{\mathrm{M}^{+}} (w)$ depends on the
energetic interactions of solution constituents, i.e., the solvent and
dissolved species, with M$^{+}$ ions, and would vanish if those interactions
were to vanish.  That is also the case for the contribution $e\phi (w)$.

 The separation imposed by Eq.~\ref{chempot} is clear on the basis of our
 mechanical understanding of electrostatics.  But from a thermodynamical 
 view it is a \emph{nonoperational}  separation \cite{KO}.

As a gauge of molecular structure and energetics of the solution in
the neighborhood of an M$^{+}$ ion, $\overline\mu_{\mathrm{M}^{+}}
(w)$ is worthy of interest.  In contrast, $\phi (w)$ reflects
distribution of charge on the surface of the system, and is of less
intrinsic interest because it is not sensitive exclusively to the local
condition of an ion of interest.  The goal is to separate
$\overline\mu_{\mathrm{M}^{+}} (w)$ from thermodynamic combinations such as $\overline\mu_{\mathrm{M}^{+}} (w)$ +
$\overline\mu_{\mathrm{X}^{-}} (w)$, to which the potentials of the
phases don't contribute, \emph{and} to separate
$\overline\mu_{\mathrm{M}^{+}} (w)$ from $e\phi (w)$.

Our notation here is deficient in one respect.
For equilibrium of two conducting phases containing ionic solutes at
low concentration, the potential difference $\Delta \phi$ =
$\phi(\gamma)$ - $\phi(\eta)$ across the phase boundary is not simply  
a property of the pair of solvents in contact with one another
\cite{lrp:jpc92,ZHOUYQ:NOTSFT}.  (Note that many materials can
be considered conductors for sufficiently long times.) From a formal
point of view of 
macroscopic thermodynamics, this follows from the requirement that the
bulk compositions of a macroscopic conductor be electrically neutral.
More physically, the ions make a Donnan-like contribution to the
difference between the electrostatic potentials of the phases to
achieve neutral bulk compositions.  This potential difference does not
progressively vanish as the concentration of the ionic solutes
decreases, so it is not operationally avoidable by confining attention
to low electrolyte concentrations.  The difference
$\Delta \phi$, whatever its source, does not contribute to the
thermodynamic combinations $\overline\mu_{\mathrm{M}^{+}} (w)$ +
$\overline\mu_{\mathrm{X}^{-}} (w)$. Therefore an accurate evaluation
of $\overline\mu_{\mathrm{M}^{+}} (w)$ on any basis would avoid
consideration of $\Delta \phi$ altogether.

Thermodynamic considerations will involve differences in these free
energies and differences $\Delta \phi$ in the electrostatic
potentials.  If a dilute gas is considered to be the coexisting conducting phase,
it is instinctive to assign the potential of that phase to be zero.  No
other considerations need be affected by this choice.  It is then
convenient to take $\phi(w)$, for example, to be the potential
relative to a dilute gas phase.

\section{Analyses Based upon Ion-Water Cluster Data}

Discussions of the difficulty of obtaining the absolute hydration free
energy $\overline \mu_{\mathrm{M}^{+}} (w)$ often begin with the
observation that thermodynamic processes for bulk phases involve
manipulation of neutral combinations of material.  It is natural,
therefore, to consider adding a single ion to a sub-macroscopic
amount of solvent, for example
\begin{eqnarray}
\mathrm{M}^+ + \left(\mathrm{H}_2\mathrm{O}\right)_n \rightarrow \mathrm{M}\left(\mathrm{H}_2\mathrm{O}\right)_n {}^+
\label{reaction}
\end{eqnarray}
for $n$ not too large.  This leads to the consideration of ion-water
cluster experiments as a potential source of information on $\overline
\mu_{\mathrm{M}^{+}} (w)$.  Coe \cite{coe:irpc01} helpfully reviews
that work.  These experimental methods permit determination of a
standard contribution to the chemical potential of the cluster on the
right side of Eq.~\ref{reaction} in a dilute gas, for cluster sizes
typically including $n$=4, 5, 6.  It is helpful to keep in mind that
these quantities can be targets of molecular simulation
\cite{LuDS:Ionsmp,lrp:jpca98} as well. 

We collect several relevant points about these quantities. A first
point is that the standard chemical potential of such a cluster
includes contributions from the kinetic motion of the ion over
the interior of the cluster. For mesoscopic clusters, this contribution may be
awkward because the volume of the cluster accessible to the ion can be ambiguous.  An extreme example of this ambiguity is the phenomenon, now well appreciated \cite{PereraL:IONSIW,%
 JORGENSENWL:LIMEOP,DANGLX:MOLSOA,SremaniakLS:ENTOFA,PereraL:STROCA,%
 MarkovichG:ThesCB,%
YehIC:PhosCc,CabarcosOM:Thescb,StuartSJ:Effphc,StuartSJ:Surcea,%
TobiasDJ:Sursha,PeslherbeGH:Cluitp},  that some ions might be localized on  the surfaces 
of small water clusters. Fortunately, the cluster sizes studied in ion-water cluster experiments considered
here are so small that even  the surface {\em vs.} interior question doesn't appear
to be a  significant problem.

A second basic point is that neither the experiment nor the
typical simulation calculations further separate
$\Delta\mu^{}_{\mathrm{M}^{+}}(w)\equiv$ $ e\phi +
\overline\mu^{}_{\mathrm{M}^{+}}(w)$ into an absolute hydration free
energy and a contribution from the potential of the phase.  This is
natural because both of these contributions derive from intermolecular
interactions, and in the case of simulations it would take a special separate calculation to
evaluate the potential of the phase separately.

Our final basic point about the separation Eq.~\ref{chempot} is that the difficulties in separating the 
various contributions to the interaction contribution to the chemical
potential Eq.~\ref{chempot} from physical data are not trivially
simpler for a particular $n$.  Though the contribution $e\phi$ is 
expected to depend on $n$, it 
will typically vanish neither for small nor for large values of $n$.

Returning to the experimental analysis, we note that the conventional
hydration free energies of ions are referenced to the case for
H$^+$(aq).  So the conventional hydration free energy for the M$^{+}$
ion is $\overline\mu^{}_{\mathrm{M}^{+}}(w) -
\overline\mu^{}_{\mathrm{H}^{+}}(w)$ and similarly that for X$^-$ is
$\overline\mu^{}_{\mathrm{X}^{-}}(w) +
\overline\mu^{}_{\mathrm{H}^{+}}(w)$.  Following Klots
\cite{klots:jpc81}, we then form the following combination and obtain:
\begin{eqnarray}
\Delta\mu^{}_{\mathrm{M}^{+}}(w) - \Delta\mu^{}_{\mathrm{X}^{+}}(w) &  =  & 2( e\phi (w)  +\overline\mu^{}_{\mathrm{H}^{+}}(w)) \nonumber \\
& + & (\overline\mu^{}_{\mathrm{M}^{+}}(w) - \overline\mu^{}_{\mathrm{H}^{+}}(w) )  \nonumber \\
& - & (\overline\mu^{}_{\mathrm{X}^{-}}(w) + \overline\mu^{}_{\mathrm{H}^{+}}(w))~.
\label{Klots}
\end{eqnarray}
The last two contributions in parenthesis on the right are
\emph{conventional} single ion hydration free energies and can be
obtained from tables.  Klots gives a plausible procedure for obtaining
the quantities on the left from the ion-cluster experimental results.
\emph{If} the quantities on the left of Eq.~\ref{Klots} were
obtainable to sufficient accuracy from cluster studies
(Eq.~\ref{reaction}), then Eq.~\ref{Klots} could be solved for $
e\phi(w) + \overline\mu^{}_{\mathrm{H}^{+}}(w)$ =
$\Delta\mu^{}_{\mathrm{H}^{+}}(w)$, corresponding to the same
standard state as for H$^+$ in compiling the conventional ion
hydration free energies.  The evidence presented
\cite{coe:irpc01,klots:jpc81} suggests that this procedure is
successful to an interesting degree, and remarkably so for small values of 
$n$.  In view of those results and the analysis from
Eq.~\ref{Klots}, we conclude that the values otained have not
separated the absolute hydration free energy
$\overline\mu^{}_{\mathrm{M}^{+}}(w)$ from the potential of the phase
in $\Delta\mu^{}_{\mathrm{H}^{+}}(w)$.

In concluding this section, we return to consider the  \emph{Donnan-like}
contribution to $\Delta\phi$ that was noted above \cite{lrp:jpc92,ZHOUYQ:NOTSFT}.   This electrostatic contribution
has its source in sorting of ions of different charges within the ion correlation length
of an  interface.   Though the source of this contribution is the ions distributed in 
interfacial regions, this  contribution typically achieves a non-zero value in the limit
of vanishing ionic contributions.  This contribution  is present in the macroscopic description Eq.~\ref{chempot}. But because the ion-water cluster experiments considered here treat clusters with only  one ion that Donnan-like ionic effect isn't present in those experimental data.

\section{Quasi-chemical Theory}

In the quasi-chemical theory \cite{lrp:apc02}, the
region around the solute of interest is partitioned into inner and
outer shell domains. A variational check of this partition exists, if
desired \cite{lrp:fpe01}.  Here we treat the inner
shell, where chemical effects are important, quantum mechanically. The
outer shell contributions have been assessed using both a dielectric
continuum model and classical molecular dynamics simulations.

The inner shell reactions pertinent to the hydration of
$\mathrm{X^{\pm}}$( = H$^+$, Li$^+$, 
Na$^+$, HO$^-$) are
\begin{eqnarray*}
\mathrm{X^{\pm} + n H_2O \rightleftharpoons X\cdot[H_2O]_n{}^{\pm}}
\end{eqnarray*}
Based on earlier work, the inner shell hydration numbers are
$n=(2,4,4,3)$ for H$^+$ \cite{lrp:jpca02}, Li$^+$ \cite{lrp:jacs00},
Na$^+$\cite{lrp:fpe01}, and HO$^-$\cite{asthagir:02},  respectively.
The free energy change for these reactions were calculated using the 
Gaussian suite of programs \cite{gaussian}. The
$\mathrm{X\cdot[H_2O]_n{}^{\pm}}$ clusters (and $\mathrm{H_2{}O}$) were 
geometry optimized in the gas phase using the B3LYP hybrid density
functional\cite{b3lyp} and the 6-31+G(d,p) basis set. Frequency
calculations confirmed a true minimum, and the zero point energies
were computed at the same level of theory. Single point energies were
calculated using the 6-311+G(2d,p) basis set. 

{\bf Dielectric continuum model:} For estimating the outer shell electrostatic 
contribution, the ChelpG method \cite{breneman:jcc90} was used to obtain 
partial atomic charges. The non-electrostatic contributions are 
expected to make negligible contributions to the hydration free energy
and are not considered further. Then with the radii set developed by
Stefanovich et al.\cite{Stefanovich:cpl95}, surface tessera were 
generated \cite{sanner}, and the hydration free energies of the
clusters were calculated using a dielectric continuum model
\cite{lenhoff:jcc90}. With this information and the binding free
energies for the chemical reactions, a primitive quasi-chemical
approximation to the excess chemical potential of X$^{\pm}$(aq) in
water is:
\begin{eqnarray}
\beta \overline \mu_{\mathrm{X}^{\pm}(w)} &\approx& - \ln \left (
\tilde{K}_n \rho_{\mathrm{H}_2\mathrm{O} }{}^n  \right)\label{eq:regrouped}
\end{eqnarray}
where $\tilde{K}_n=K_n{}^{(0)}\exp\left[{-\beta
\left(\overline\mu_{\mathrm{X}(\mathrm{H}_2\mathrm{O})_n{}^{\pm}}(w)-n 
\overline\mu_{\mathrm{H}_2\mathrm{O} }(w)\right)}\right]$. $K_n{}^{(0)}$ is
the equilibrium constant for the reaction in an ideal gas state, with
$n$ of Eq.~\ref{eq:regrouped} the hydration number of the most stable
inner shell cluster, and $\mathrm{\beta=1/k_\mathrm{B}T}$.  The
density factor $\mathrm{\rho_{H_2O}}$ appearing in
eq.~\ref{eq:regrouped} reflects the actual density of liquid water and
its effect is accounted for by including a replacement contribution of
$\mathrm{-n k_\mathrm{B}T \ln (1354)}$.  A detailed statement on standard
states and this replacement contribution can be found in
Grabowski~et~al.~\cite{lrp:jpca02}. Relevant energies are collected in
TABLE~\ref{tb:ener}.  
\begin{table}[h]
\caption{Electronic energy (a.u.), thermal corrections (a.u.) to the
free energy, and electrostatic contributions to the excess chemical
potential (kcal/mole) using dielectric continuum (DC) approximation
and molecular dynamics (with TIP3P and SPC/E water models). The
statistical uncertainties in the molecular dynamics values 
are of the order of 1.5~kcal/mole (Fig.~\ref{fg:md}). The
partial charges are obtained at B3LYP/6-311+G(2d,p).}\label{tb:ener}  
\begin{tabular}{lrrrrr}\hline
 \multicolumn{3}{c}{} & \multicolumn{3}{c}{$\mu^{ex}$} \\
  & \multicolumn{1}{c}{E} & \multicolumn{1}{c}{G$_{corr}$}  & DC & TIP3P & SPC/E \\ \hline
H$_2$O & -76.45951 & 0.00298 & -7.7 & -6.6 & -6.5 \\
H$^+$ & 0 & -0.01 & -- & -- & -- \\
H[H$_2$O]$_2{}^+$ & -153.24860 & 0.03253 & -77.5 & -66.0 & -64.5 \\
HO$^-$ & -75.82779 & -0.00771 & -- & -- & -- \\
HO[H$_2$O]$_3{}^-$ & -305.32036 & 0.04705 & -72.7 & -85.2 & -87.6 \\
Li$^+$ & -7.28492 & -0.012748 & -- & -- & -- \\
Li[H$_2$O]$_4{}^+$ & -313.29314 & 0.05758 & -64.0 & -54.2 & -51.4 \\
Na$^+$ & -162.08757 & -0.014429 & -- & -- & -- \\
Na[H$_2$O]$_4{}^+$ & -468.05512 & 0.05095 & -62.0 & -51.5 & -49.8 \\ \hline
\end{tabular}
\end{table}
For the dielectric continuum calculations, an observation volume of
radius 1.9~{\AA} and 2.3~{\AA} was centered on the Li$^+$ and Na$^+$
ions, respectively. For H$^+$ and HO$^+$ these radii were 1.172~{\AA}
and 1.56~{\AA} respectively. The calculated values change only
modestly to increases in these values. 

{\bf Classical molecular dynamics:} The electrostatic contribution to
the hydration (excess) free energy is given by 
\begin{eqnarray}
\overline\mu = \int_0^q \langle \psi \rangle_\lambda d\lambda,
\end{eqnarray}
where $\lambda$ is the coupling parameter which switches the solute
charge from 0 to q. $\langle\psi\rangle_\lambda$ is the ensemble-averaged
potential on the ion at a particular charge state, $\lambda$. Gauss-Legendre
quadratures \cite{hummer:jcp96} provides a facile route to estimate
this integral. In particular, treating $\langle\psi\rangle_\lambda$ as a third
degree polynomial, the following two-point formula
\begin{eqnarray}
\overline\mu \approx \frac{q}{2} ( \langle \psi \rangle_{\lambda_+} +
\langle \psi \rangle_{\lambda_-} )~, 
\end{eqnarray}
where $\lambda_{\pm} = q(1/2 {\pm} 1/\sqrt{12})$, is
exact to 4$^\mathrm{th}$ order in perturbation theory
\cite{hummer:jcp96}. Trial calculations using a purely linear response
approximation (2$^{nd}$ order in perturbation theory) deviate modestly
from the above approximation, and hence for greater realism we
employed the above form. For clarity the above expressions pertain to
an atomic ion, but are easily generalized to complex ions with
distributed charges.  

Since the solute is largely buried within the first shell water
molecules, there is no need for developing accurate parameters for
those. This is true for the cations. For HO$^-$, the hydroxyl hydrogen
is largely free \cite{asthagir:02}. But the charge on this hydrogen is
small, being about half the charge of the hydrogen atoms 
in the classical water models \cite{asthagir:02}. 
Based on these observations, the van~der~Waals parameters for Li$^+$ and Na$^+$
were obtained from \cite{aqvistli} and \cite{rouxna}, respectively.
The H and O atoms in the cluster were assigned the same van~der~Waals
parameters as those in classical water model, of which both
TIP3P\cite{tip31,tip32} and SPC/E\cite{spce} were used. The charge
distribution on these clusters were the ChelpG charges. The 
clusters were simulated in simulation cells of different
sizes to estimate ion finite-size effects, and included 32, 64, 216
and 306 water molecules in addition to the cluster of interest.

The simulations were all performed with the NAMD \cite{namd} code
which uses  the particle mesh Ewald method to treat long range
electrostatic interactions. For each $\lambda$, the simulation
consisted of 200~ps of equilibration with velocity scaling every
250~fs, followed by another 200~ps of equilibration with velocity
scaling every 1~fs. The water geometry was constrained using 
SHAKE. The cluster was held fixed. For the 32 and 64 water molecule
boxes, a production run of 375~ps was conducted. For the 216 and 306
water molecule boxes, the production run lasted 150~ps. Frames were
stored every 5~fs. Then using programs developed in-house, the Ewald 
potential at the solute charge sites were calculated, and statistical
uncertainties were estimated by block-averaging in block sizes of
7.5~ps, corresponding roughly to the dielectric relaxation of pure
water \cite{neumann:jcp86}. Then the free energy was assembled.

Electrostatic finite size corrections were applied as described in
Hummer et al.\ \cite{lrp:jpca98}. As in Hummer et al.\
\cite{lrp:jpca98}, the real-space screening factor was $\eta=5.6/L$,
where $L$ is the box length. A cutoff of $k^2 \leq 38(2\pi / L)^2$ was
applied in Fourier space, resulting in $2 \times 510$ {\bf k}
vectors. A cutoff of L/2 was applied for the real-space electrostatic
interactions.  

To check the accuracy of our code, we simulated some earlier
results for ion hydration, including the charging free energy of
imidazolium \cite{lrp:jpca98}. Hummer et al.\  \cite{lrp:jpca98}
studied imidazolium charging using Monte Carlo techniques in boxes
ranging from 16 water molecules to 512 water molecules. In the 16,
64, and 512 water molecule boxes, they find that the average
interaction energy at full charge is  
$-119.7\pm0.24$~kcal/mole, $-113.3\pm0.36$~kcal/mole,  and
$-110.3\pm0.4$~kcal/mole, respectively. We find
$-118.9\pm1.6$~kcal/mole, $-113.6\pm1.3$~kcal/mole, and
$-111.3\pm2.2$~kcal/mole, respectively. Given the different
methodologies this agreement is good.  

\begin{figure}[h]
\begin{center}
\includegraphics[width=3.00in]{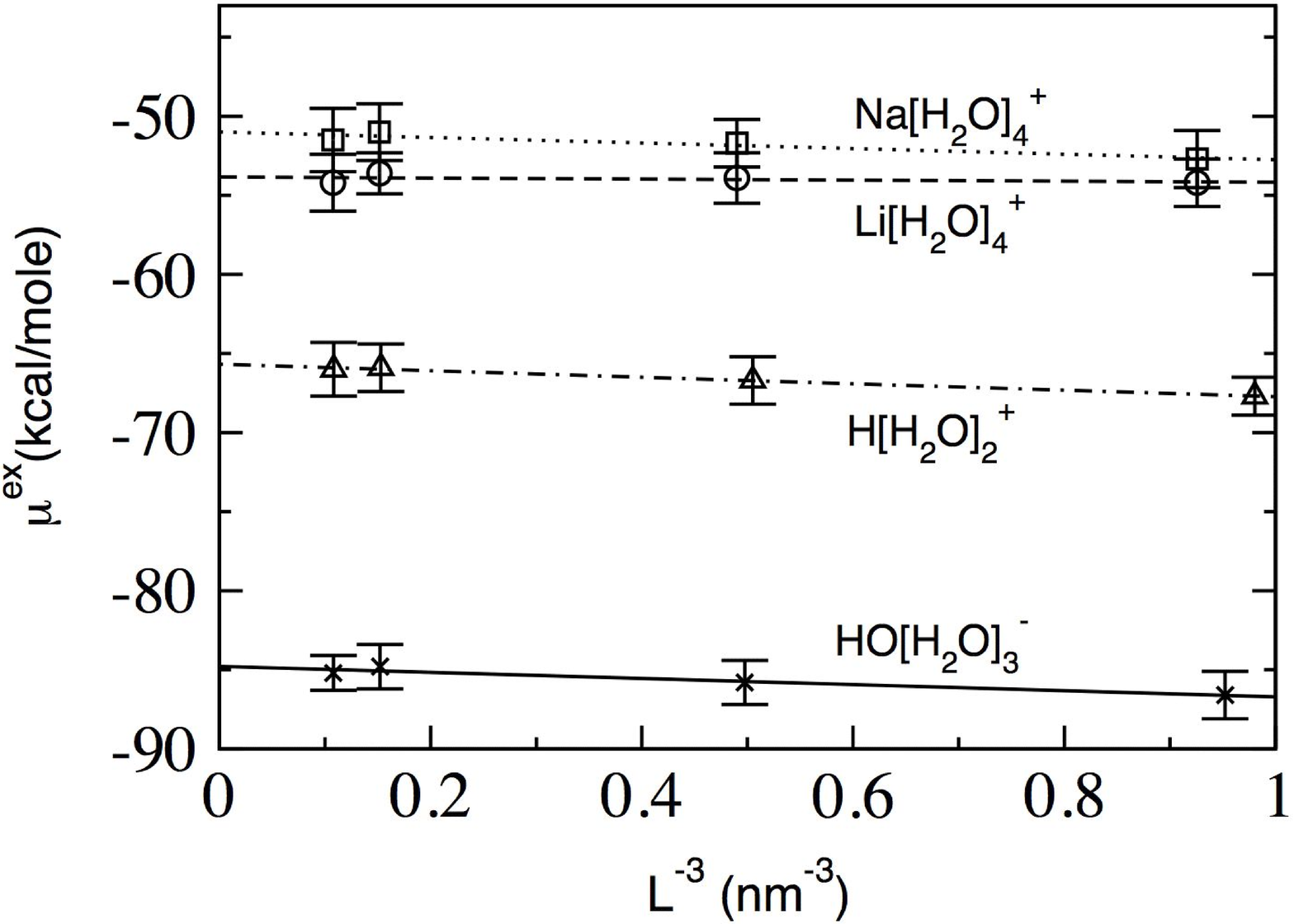}
\end{center}
\caption{Electrostatic contribution to the excess chemical potential
of the clusters using the TIP3P water model. $\mathrm{L}$ is the box size and
from left to right, the boxes contain 306, 216, 64, and 32 water
molecules, respectively. The box volume is adjusted to give  a density
of 1~gm/cc including the water molecules part of the cluster. The
ions are given a partial specific volume of zero (0), but this choice
is more for convenience and a more appropriate negative value would
not change the results substantially.}\label{fg:md} 
\end{figure}

Figure~\ref{fg:md} shows that for the clusters under consideration,
ion finite size effects are quite modest. The values in
table~\ref{tb:ener} using the TIP3P water model are obtained from
fig.~\ref{fg:md}. The least square line fits are weighted by the statistical
error.  Since the ion finite size effect is small, the TIP3P and SPC/E values 
(TABLE~\ref{tb:ener}) were all obtained with the box of 306 water
molecules.

\section{Results}

The hydration free energies obtained with the
three different calcualtions of the outer-sphere contributions are 
given in TABLE~\ref{tb:hyd}.
\begin{table}[h]
\caption{Excess chemical potentials (kcal/mole) of the solutes
obtained with outer-sphere contributions based on the dielectric
continuum (DC) and classical molecular dynamics approaches. The
experimental values suggested by Coe et al.~\cite{coe:jpca98}  
pertaining to the transfer of the solute from 1~atm (ideal gas) to 1~M
(ideally diluted solute) is included after adjustment for 1~M (ideal
gas) to 1~M (ideally diluted solute) transfer. }\label{tb:hyd}
\begin{center}
\begin{tabular}{lrrrr} \hline
  &  DC & TIP3P & SPC/E & Expt \\ \hline
H$^+$ & -254.6 & -245.3 & -244.0 & -265.9 \\
Li$^+$ & -120.5 & -115.1 & -112.7 & -128.4 \\
Na$^+$ & -96.1 & -90.0 & -88.7 & -103.2 \\
HO$^-$ & -105.3 & -121.1 & -123.8 & -104.9 \\ \hline
\end{tabular}
\end{center}
\label{II}
\end{table}
The agreement between our absolute hydration free
energies and Coe and coworkers' \cite{coe:jpca98,coe:irpc01} values is
poor for the molecular dynamics results, and fairs 
only somewhat better for the 
dielectric continuum results. The computed hydration free energies for neutral
ion pairs, however, are in markedly improved agreement with the experimentally 
suggested values table~\ref{tb:neutral}, particularly for the molecular 
dynamics vales which are in error by about 2-3 kcal/mole compared to the
dielcetric continuum predictions which err by 8-11 kcal/mole.
\begin{table}[h]
\caption{Solvation free energy of neutral ion pairs (kcal/mole). The
solutes are transferred from 1~M (ideal gas) to 1~M (ideally diluted
solute). Experimental numbers are from table~\ref{tb:hyd}
above. $\dag$ For water, based on the experimental gas phase free
energy of dissociation (384.1~kcal/mole), the known $\mathrm{pK_w}$ of
water (15.7) and the known hydration free energy of water
($-6.3$~kcal/mole), we calculate a value of $-366.6$~kcal/mole for the HOH
pair. Klots \cite{klots:jpc81} quotes a value of
$-368$~kcal/mole for this quantity.}\label{tb:neutral}  
\begin{center}
\begin{tabular}{lrrrr} \hline
  &  DC & TIP3P & SPC/E & Expt \\ \hline
HOH & -359.9  & -366.4 & -367.8 & -370.7$\dag$ \\
LiOH & -225.8 & -236.2 & -236.5 & -233.3 \\
NaOH & -201.4 & -211.1 & -212.5 & -208.1 \\ \hline
\end{tabular}
\end{center}
\end{table}

{\bf $\mathrm{pK_w}$ of water:} A further test of the pair hydration
free energy for H$^+$ and HO$^-$ is to compute the $\mathrm{pK_w}$, a
quantity of immense interest in solution chemistry, especially protein
biochemistry. 

Earlier, Haymet and coworkers \cite{haymet}, Guissani and
coworkers \cite{guissani} and Tawa and Pratt \cite{lrp:jacs95} considered
the ionization of water. Tawa and Pratt \cite{lrp:jacs95}  went further, 
describing this ionization at different thermodynamic states.
These are important works, but are not {\em ab initio \/} descriptions of dissociation. 

The reaction under consideration is the ionization of water (in water
solvent): 
\begin{eqnarray*}
\mathrm{H_2O \rightleftharpoons H^+ + HO^-}
\end{eqnarray*}
The equilibrium constant for the reaction can be written as
\begin{eqnarray}
\mathrm{K_w = \frac {q_{H^+}q_{HO^-}}{q_{HOH}} (\rho/\rho_o{})}
\end{eqnarray}
Here $\mathrm{\rho_o}$=1~M is the reference concentration 
for H$^+$ and HO$^-$. $\rho$=55~M is the reference concentration of
water. $\mathrm{q_i = q_i{}^{ig}
\langle\langle \exp^{-\beta \Delta U} \rangle\rangle_o}$ is the
partition function. $\mathrm{q_i{}^{ig}}$ is the ideal gas partition
function, and $\mathrm{\langle\langle \exp^{-\beta \Delta U}
\rangle\rangle_o}$ is the Widom factor. $\mathrm{\Delta U}$ is the
interaction energy of the solute with the solvent
\cite{lrp:jpcb02}. $\mathrm{-RT \ln \langle\langle \exp^{-\beta \Delta
U} \rangle\rangle_o }$ is precisely the excess chemical potential 
obtained above using the quasi-chemical approach.

For the ionization reaction in gas phase, the free energy change is
computed (using TABLE~\ref{tb:ener}) to be 383.4~kcal/mole, which
compares favorably with 384.1~kcal/mole estimated experimentally
\cite{schulz:jcp82,bartmess}. The calculated $\mathrm{pK_w}$'s are
21.1, 15.6 and 14.5, respectively, for the outer-sphere contributions
using dielectric continuum, TIP3P water, and SPC/E water. The agreement 
between the molecular dynamics results and the experimental 
value of 15.7 \cite{pearson:jacs86} is excellent. 

\section{Discussion}

From a molecular standpoint, the quantity of most interest is
$\overline\mu_{\mathrm{X}}$, the absolute hydration free energy. This absolute
quantity is understood to be relative to the average potential of the 
phase, or equivalently to the value of that potential of the  phase being taken as zero.

In the Ewald summation method  there is no interface, and  the mean potential in the cell is zero. Thus, use of  Ewald potential to compute charging free energies naturally conforms to
computation of the absolute hydration free  energy.

In a cluster, however, the reference potential is not
zero. Simulations of the vapor-water interface by Sokhan and
Tildesley\cite{tildesley:mp97} suggests that the potential in bulk
SPC/E water is -12.7~kcal/mole-e relative to the vapor value of 0
(zero). Adding this value to our SPC/E estimate we obtain
-256.7~kcal/mole for the hydration free energy of the proton in the
presence of the potential of the phase. This greatly improves the
agreement with the value of the value of -265.9~kcal/mole obtained by
Coe and coworkers \cite{coe:cpl94,coe:jpca98,coe:irpc01} and
-264.3~kcal/mole obtained by Klots \cite{klots:jpc81}. Consistent with
this but probably more significantly, the consensus results of
TABLE~\ref{II} suggest a value that is negative and somewhat greater
than 10~kcal/mole-e in magnitude.

A classic alternative to the ion-cluster experimental studies for
obtaining the absolute hydration free energies of ions is the
tetraphenyl arsonium tetraphenyl borate (TATB) hypothesis
\cite{SchurhammerR:HydiTw}.  The molecular intuition is that these
oppositely charged ions have the same non-ionic interactions with any
solvent molecules.  If it were precisely true that the absolute
hydration free energy of these two solutes were precisely the same,
then the Donnan-like effects discussed above would imply $\Delta\phi$
= 0 at equilibrium of this salt between two coexisting fluids.  If
true, this would be a satisfactory solution of the present problem.
But because the values obtained that way would not include a contribution
from the potential of the phase, those values would be different from
values extracted from the ion-cluster studies analyzed here.  The
computational testing of that TATB hypothesis has lead to energetic
uncertainties of nearly the same size as the energetic contributions
of issue here \cite{SchurhammerR:HydiTw}.  Therefore, the TATB
hypothesis must be currently viewed as not satisfactorily proven.  One
important consideration is the following: because of the particular
electrical asymmetry of water molecules, positively and negatively
charged ions with exactly the same non-ionic interactions with water
molecules will not have the same absolute hydration free energies.  It
is commonly observed in simulation that anions are better hydrated
than cations of the same non-ionic interactions
\cite{Friedman:73,lrp:jpc96,AshbaughHS:Conmam}; an intuitive view of
that phenomenon is that water molecule hydrogen atoms, carrying some
positive charge, are able to approach anionic solutes closely and
this is different for the corresponding cationic solutes.

 Our computed hydration free energy estimates, especially for ion pairs, is in good agreement with experiments. This agreement should not obscure the approximations made in applying a
rigorous theory to practical calculations. These calculations might be improved in several
 respects.  For example, in calculating the excess free energy of the cluster and the water
ligand, we neglected packing effects and dispersion interactions. The
tacit assumption in the analysis is that in composing $K_n$
(eq.~\ref{eq:regrouped}), errors in the cluster and the $n$ water
ligands balance out. For ionic species this appears to be a good
assumption, but it does fail for hydration of non-polar solutes 
\cite{lrp:bc02}. We foresee refining these aspects of the calculation
in future studies. 

The neglect of  packing contributions is probably more easily remedied; see the discussion
 \cite{Pratt:ES:99}. Packing contributions are expected to be only a few percent of  the chemical and electrostatic effects included already.  But those packing contributions are expected to be positive.  Those contributions would probably improve the agreement for LiOH and NaOH in
TABLE~\ref{tb:neutral}.  On the otherhand, several additional effects of  comparable size would have to be included at the same time.  For example, the effects of anharmonicity on the thermal 
 motion of the clusters is a comparable worry.

\section{Conclusions}

We find that the absolute hydration free energy,
$\overline\mu_{\mathrm{H^+}}(w)$, is somewhat smaller than the
acknowledged $-251$ to $-264$~kcal/mole span of numbers.  The present
analysis suggests that the more negative of these values include the
potential of the phase, a contribution distinct from the absolute
hydration free energy.

Calculations on ion-water clusters can be compared properly with
experiments on ion-water clusters.  Those results and comparisons can
test the adequacy of such calculations.  But they don't exclusively
test the description of hydration of the ion; part of those results
derives from the surface structure of the cluster and that
contribution does not vanish for large clusters.

This  study reinforces the idea that strenuous calculation of the absolute hydration free energy, $\overline\mu_{\mathrm{M^+}}(w)$, for a favorable case such as M=Li could probably reduce
the  computational errors to less than the  present experimental uncertainties.  This would permit
a more precise alignment of available tables  of single ion absolute hydration free energies.

\section{Acknowledgements}

The work at Los Alamos was supported by the US Department of Energy,
contract W-7405-ENG-36, under the LDRD program at Los
Alamos. LA-UR-03-1731.

\end{document}